\documentclass[pss]{wiley2sp}
\usepackage{amsmath}
\newcommand{\ket}[1]{\left| #1 \right \rangle } 
\newcommand{\bra}[1]{\left \langle #1 \right |} 
\newcommand{\bsup}{\begin{subequations}}
\newcommand{\esup}{\end{subequations}}
\newcommand{\ulc}{\hspace{0.03cm}\underline{\phantom{i}}\hspace{-0.135cm}\hat{c}}
\newcommand{\ulPsi}{\hspace{0.03cm}\underline{\phantom{i}}\hspace{-0.14cm}\Psi}
\newcommand{\ulm}{\hspace{0.03cm}\underline{\phantom{n}}\hspace{-0.252cm}\hat{m}}
\newcommand{\ulH}{\hspace{0.03cm}\underline{\phantom{J}}\hspace{-0.24cm}\hat{H}}
\newcommand{\ulHcal}{\hspace{0.03cm}\underline{\phantom{I}}\hspace{-0.23cm}\mathcal{H}}
\newcommand{\ulO}{\hspace{0.03cm}\underline{\phantom{I}}\hspace{-0.23cm}\hat{O}}
\newcommand{\ulI}{\hspace{0.03cm}\underline{\phantom{i}}\hspace{-0.12cm}I}
\newcommand{\ulGamma}{\hspace{0.03cm}\underline{\phantom{i}}\hspace{-0.13cm}\Gamma}
\newcommand{\hc}{\hat{c}^{\phantom{\dagger}}}
\newcommand{\hcd}{\hat{c}^{\dagger}}
\newcommand{\langind}[1]{\hspace{-0.1cm}{\phantom{\rangle}}_{#1}
\hspace{-0.05cm} \langle\hspace{0.03cm} }  
\newcommand{\rangind}[1]{\rangle\hspace{-0.2cm}{\phantom{\rangle}}_{#1}
\hspace{0.05cm}}
\begin{document}
\title{A slave-boson mean-field theory for general multi-band Hubbard models}
\titlerunning{Slave-boson  theory for multi-band models }
\author{J. B\"unemann}
\authorrunning{J. B\"unemann}
\mail{e-mail
  \textsf{joerg.buenemann@physik.uni-marburg.de}, Phone:
  +49-6421-2824191}
\institute{Fachbereich Physik,
Philipps Universit\"at, Renthof 6, 35032  Marburg, Germany}
\received{XXXX, revised XXXX, accepted XXXX} 
\published{XXXX} 

\abstract{
\abstcol{We introduce a new slave-boson mean-field theory which allows the 
 investigation of general multi-band Hubbard models. Unlike earlier 
 attempts of such a generalisation, in our approach  the quantum-mechanical 
problem    is exactly  reformulated in an extended Hilbert space}{ of  Fermions and 
Bosons before a mean-field approximation is applied.
  Systems with superconducting order parameters are naturally included in 
our formalism. Our ground-state energy functional agrees with 
the corresponding quantity derived within the Gutzwiller theory.}}

\maketitle
\section{Introduction}
For the investigation of correlated electron systems in two or three dimensions
 only few theoretical approaches exist which lead to sensible results 
 for small, medium, as well as strong Coulomb interaction parameters. A widely
 used approach which fulfils this criterion 
 is the slave-boson mean-field theory. 
It has been introduced by Kotliar and Ruckenstein 
 for an investigation of the single band Hubbard model 
\cite{kotliar1986}. The approach is
 based on two main ideas: 
\begin{itemize}
\item[i)]The original quantum-mechanical problem, described by a 
 fermionic Hamiltonian $\hat{H}$,
 is reformulated by introducing 
 bosonic degrees of freedom at each lattice site $i$ of the system. As long 
 as certain constraints are exactly fulfilled, the resulting new
 Hamiltonian $\ulH$ in the enlarged bosonic-fermionic 
Hilbert space  is mathematically equivalent to $\hat{H}$.  
 \item[ii)] The Hamiltonian  $\ulH$ is investigated by means of a mean 
field theory or, in the 
 language of functional integrals, by a saddle-point approximation.      
\end{itemize}
Unfortunately, there is an infinite number of Hamiltonians $\ulH$, 
 equivalent to $\hat{H}$, which may all lead to different results 
on mean-field level. Therefore, the approach requires a sophisticated 
 guess in order to find the `right'  Hamiltonian which yields
 sensible mean-field results.
The choice, made by Kotliar and Ruckenstein, leads to the same 
ground-state energy functional as it is found by an 
exact evaluation of Gutzwiller wave-functions in the limit 
of infinite spatial dimensions \cite{gebhard1990,gebhard1991}.   
     
For the investigation of  real materials one usually has to take 
 into account the  multi-orbital electronic structure of such systems. 
This requires 
 the study of multi-band Hubbard models. While the generalisation of the 
 Gutzwiller theory for multi-band models is rather 
straightforward even for systems with superconducting ground states~\cite{buenemann1998,buenemann2005,buenemann2005b}, 
the very same generalisation 
of the  slave-boson theory turned out to be difficult. Only recently 
such a generalised slave-boson scheme
 has been introduced for the treatment of general multi-band 
models~\cite{lechermann2007}. The ground-state energy functional, 
 derived in that work, has been shown to agree with the 
Gutzwiller functional \cite{buenemann2007c}.
In a more recent work \cite{isidori2009}, the scheme introduced in 
 \cite{lechermann2007} was generalised in order to study 
superconducting systems. For such systems, however, the ground-state energy 
functional derived in  \cite{isidori2009} seems not to agree with the 
corresponding  Gutzwiller functional~\cite{buenemann2005}.

In the derivations of 
Refs.~\cite{lechermann2007,isidori2009} 
there is a fundamental shortcoming. While in the single-band case 
the derivation of the slave-boson theory is exact up to the point 
where the mean-field approximation is applied, 
 the multi-band  derivation of Ref.~\cite{lechermann2007} 
requires additional approximations already on the level of the operator 
equations. It is the main purpose of this work to introduce an alternative 
  slave-boson scheme for multi-band models which avoids these additional 
approximations and is exact 
apart from the final mean-field treatment. 
In addition, our new approach automatically
 covers systems with superconducting order parameters.

The presentation is organised as follows: In section~\ref{chap2}, we introduce 
 the general multi-band models that are investigated in this work. We briefly 
  summarise the results of the Gutzwiller theory for multi-band 
 Hubbard models in section ~\ref{chap2b}.
 A reminder of the Kotliar-Ruckenstein theory for the one-band Hubbard model 
 is given in  section \ref{chap3a}. Previous attempts to generalise this
 approach 
 are discussed  in section~\ref{chap3b}. In section~\ref{chap5},
  we derive our new slave-boson theory for the investigation of 
general multi-band models. 
A summary closes our 
 presentation in section~\ref{chap6}.

\section{Model Hamiltonians}\label{chap2}
We study the general class of multi-band Hubbard models
\begin{equation}\label{h2}
\hat{H}=\sum_{i\neq j} \sum_{\sigma,\sigma'}
t^{\sigma,\sigma'}_{i,j} \hcd_{i,\sigma}\hc_{j,\sigma'}+
\sum_i \hat{H}_{i,{\rm loc}}
\end{equation}
where we introduced the combined, spin-orbital indices $\sigma$, 
 hopping parameters $t^{\sigma,\sigma'}_{i,j}$, and  local 
 Hamiltonians 
\begin{eqnarray}\label{4.10a}
\hat{H}_{i;{\rm loc}}&=&\sum_{\sigma_1,\sigma_2}\varepsilon_{i;\sigma_1,\sigma_2}
\hcd_{i,\sigma_1} \hc_{i,\sigma_2}\\\nonumber
&&+\sum_{\sigma_1,\sigma_2,\sigma_3,\sigma_4}
U_i^{\sigma_1,\sigma_2,\sigma_3,\sigma_4}
\hcd_{i,\sigma_1} \hcd_{i,\sigma_2}\hc_{i,\sigma_3} \hc_{i,\sigma_4}\;
\end{eqnarray}
for each lattice site $i$. The Hamiltonian (\ref{4.10a}) is determined 
 by the orbital-dependent on-site energies  $\varepsilon_{i;\sigma_1,\sigma_2}$
 and by the 
two-particle Coulomb interaction $U_i^{\sigma_1,\sigma_2,\sigma_3,\sigma_4}$.
 It contains $2N$ local spin-orbit
 states $\sigma$, which we assume to be ordered in some arbitrary 
 way, $\sigma= 1,\ldots,2N$. Here, $N$ is the number of orbitals
 per lattice site.
In order to set up a proper basis of the local Hilbert space, 
 we introduce the following notations for the $2^{2N}$ possible 
configurations:
\newline
i) An atomic configuration $I$ is characterised by the electron 
occupation of the orbitals,
\begin{eqnarray}\label{4.20a}
\qquad
 I&\in&  \{\emptyset;
(1),\ldots,(2N);
(1,2),\ldots,(2,3),\\\nonumber
&&\ldots (2N-1,2N);
\ldots;
(1,\ldots,2N)
\}\;, 
\end{eqnarray}
where the elements in each set $I=(\sigma_1,\sigma_2,\ldots)$ 
are ordered, i.e., it is $\sigma_1<\sigma_2<\ldots$. The symbol
$\emptyset$ in~(\ref{4.20a}) means that the site is empty.
In general, we interpret the indices $I$ as sets in the usual 
 mathematical sense.
For example, in the atomic configuration
$I\backslash I'$ 
only those orbitals in $I$ that are not in $I'$ 
are occupied.
The complement of $I$ is 
$\overline{I}\equiv(1,2,\ldots,2N)\backslash I$,
i.e., in the atomic configuration $\overline{I}$ all orbitals but those
in $I$ are occupied. \newline 
ii) The absolute value $|I|$ of a configuration 
is the number of elements in it, i.e.,
\begin{eqnarray}\label{4.25a}
\qquad
&&|\emptyset|=0;|(\sigma_1)|=1;\\\nonumber
&&|(\sigma_1,\sigma_2)|=2;\ldots; |(1,\ldots,2N)|=2N
\;.  
\end{eqnarray}\newline
iii) A state with a specific configuration $I$ is given as 
\begin{equation}\label{4.30a}
\ket{I}=\hat{C}_{I}^{\dagger}\ket{0}\equiv\prod_{\sigma \in I}\hcd_{\sigma}\ket{0}=
\hcd_{\sigma_1}\dots\hcd_{\sigma_{|I|}}\ket{0}\;,
\end{equation}
where the operators $\hcd_{\sigma}$ are in ascending order, i.e., it is 
 $\sigma_1<\sigma_2\ldots<\sigma_{|I|}$. 
Products of annihilation operators, such as
\begin{equation}\label{4.35a}
\hat{C}_{I}^{}\equiv\prod_{\sigma\in I}\hc_{\sigma}=\hc_{\sigma_1}\dots\hc_{\sigma_{|I|}},
\end{equation}
will always be  placed in descending order, i.e., with 
$\sigma_1>\sigma_2\ldots>\sigma_{|I|}$. Note that we have introduced  the operators
 $\hat{C}_{I}^{\dagger}$ and  $\hat{C}_{I}^{}$ just as convenient abbreviations.
 They must not be misinterpreted as 
  fermionic creation or annihilation operators.\newline
iv) The operator $\hat{m}_{I,I'}\equiv \ket{I}\bra{I'}$ 
describes the transfer
 between configurations $I'$ and $I$. It can be written as  
 \begin{equation}\label{4.50a}
\hat{m}_{I,I'}=
\hat{C}_{I}^{\dagger}
\hat{C}_{I'}^{}
\prod_{\sigma''\in J}(1-\hat{n}_{\sigma''})
\end{equation}
where $J\equiv \overline{I\cup I'}$. A special case,
  which derives from~(\ref{4.50a}), is the occupation operator
\begin{equation}\label{4.52a}
\hat{m}_{I}\equiv  \ket{I}\bra{I}=\prod_{\sigma\in I}\hat{n}_{\sigma}
\prod_{\sigma'\in \bar{I}}(1-\hat{n}_{\sigma'})\;.
\end{equation}

The states $\ket{I}$ form a basis of the atomic Hilbert space. Therefore,
 any other basis $ |\Gamma  \rangle $ of the 
atomic Hilbert space can be written  as
\begin{equation}\label{4.60a}
|\Gamma\rangle =\sum_{I}T_{I,\Gamma}\ket{I}
\end{equation}
with coefficients $T_{I,\Gamma}$. With such a 
 general basis, the atomic Hamiltonian has the form
\bsup
\begin{eqnarray}\label{345}
\qquad
\hat{H}_{i,{\rm loc}}&=&\sum_{\Gamma,\Gamma'}
E_{i;\Gamma,\Gamma'}\hat{m}_{i;\Gamma,\Gamma'}\;,\\
\hat{m}_{i;\Gamma,\Gamma'}&\equiv&
|  \Gamma\rangind{i}
\langind{i} \Gamma'| \;.
\end{eqnarray}
\esup
  
In case that we deal with only one orbital per lattice site, the 
 Hamiltonian  (\ref{h2}) reads
\begin{equation}\label{5.20}
\hat{H}_{1{\rm B}}=\sum_{i,j} \sum^2_{\sigma=1}
t_{i,j} \hcd_{i,\sigma}\hc_{j,\sigma}+
\sum_iU_i \hat{m}_{i;12}\;.  
\end{equation}
Here, the indices $\sigma=1,2$ represent the two possible spin directions 
 and $\hat{m}_{i;12}=\hat{n}_{i,1}\hat{n}_{i,2}$ with 
 $\hat{n}_{i,\sigma}\equiv \hcd_{i,\sigma} \hc_{i,\sigma} $ 
is the `double-occupancy operator' on lattice site $i$. 

\section{The Gutzwiller Energy Functional for Multi-Band Systems}\label{chap2b}
For later comparison with the slave-boson mean-field theories,
  we briefly summarise the results of the Gutz\-willer theory for multi-band 
 Hubbard models. For all technical details, we refer the reader to Refs.
 \cite{buenemann1998,buenemann2005}. 

 Multi-band Gutzwiller wave-function have the form
\begin{equation}\label{1.3}
|\Psi_{\rm G}\rangle=\hat{P}_{\rm G}|\Psi_0\rangle=\prod_{i}\hat{P}_{i}|\Psi_0\rangle\;,
\end{equation}
where $|\Psi_0\rangle$ is a normalized single-particle product state and the 
local Gutzwiller correlator is defined as 
\begin{equation}\label{1.4}
\hat{P}_{i}=\sum_{\Gamma}\lambda^{(i)}_{\Gamma}
|\Gamma \rangle_{i} {}_{i}\langle \Gamma |\;.
 \end{equation}
 Note that, instead of (\ref{1.4}), one can also work with a
 non-diagonal variational parameter matrix $\lambda^{(i)}_{\Gamma,\Gamma^{\prime}}$,
\begin{equation}\label{1.4b}
\hat{P}_{i}=\sum_{\Gamma,\Gamma^{\prime}}\lambda^{(i)}_{\Gamma,\Gamma^{\prime}}
|\Gamma \rangle_{i} {}_{i}\langle \Gamma^{\prime} |\;.
 \end{equation}
 Since we work with an arbitrary atomic basis $|\Gamma \rangle_{i}$, however, 
 both correlation operators (\ref{1.4}) and (\ref{1.4b}) define the same 
variational space. 
In the following, we summarise the 
 main results for a correlation operator of the form (\ref{1.4}). 

In general, the local uncorrelated density matrix 
\begin{equation}
C_{i;\sigma,\sigma'}=\langle \hcd_{i,\sigma}\hc_{i,\sigma'}   \rangle_{\Psi_0}
\end{equation}
 is non-diagonal with respect to $\sigma,\sigma'$. In superconducting systems,
 it additionally exhibits anomalous elements such as 
$\langle \hcd_{i,\sigma_1}\hcd_{i,\sigma_2} \rangle_{\Psi_0}$. By means of 
 a unitary transformation (i.e., a Bogoliubov transformation for superconductors)
 one always finds a local basis with a diagonal density matrix 
 and vanishing anomalous elements. In the following, we only work with 
 such a local basis, since it simplifies the results for the variational energy.
  Nonetheless, we use the notations which we introduced in the previous section.  
 For superconducting systems, this means that the single-particle Hamiltonian
 in (\ref{h2})
 has the more complicated form
 \begin{eqnarray} 
\qquad
 \hat{H}_0=  \sum_{i\neq j}\sum_{\sigma,\sigma'}&&\left[
t^{1;\sigma,\sigma'}_{i,j} \hcd_{i,\sigma}\hc_{j,\sigma'}
+t^{2;\sigma,\sigma'}_{i,j} \hc_{i,\sigma}\hcd_{j,\sigma'}\right.\\\nonumber
&&\left.t^{3;\sigma,\sigma'}_{i,j} \hcd_{i,\sigma}\hcd_{j,\sigma'}
+t^{4;\sigma,\sigma'}_{i,j} \hc_{i,\sigma}\hc_{j,\sigma'}\right]\,.
 \end{eqnarray}
   Note that for our `orbital' basis $\ket{\sigma}$ the expectation value of the operator 
 (\ref{4.50a}) with respect to $\ket{\Psi_0}$ is given as
\bsup\label{78965}
\begin{eqnarray}
\qquad
\langle \hat{m}_{i;I,I'}  \rangle_{\Psi_0}&=&\delta_{I,I'}m^0_{i;I'}\\
m^0_{i;I'}&=&\prod_{\sigma \in I}n^0_{i;\sigma}\prod_{\sigma \notin I}(1-n^0_{i;\sigma})
\end{eqnarray}
where 
\begin{equation}
n^0_{i;\sigma}\equiv \langle \hcd_{i,\sigma} \hc_{i,\sigma}   \rangle_{\Psi_0}\;.
\end{equation}
\esup

The variational parameters $\lambda^{(i)}_{\Gamma}$ 
need to obey certain constraints, which naturally arise in 
the evaluation in infinite dimensions \cite{buenemann1998,buenemann2005}. 
These are
\bsup\label{2786}
\begin{eqnarray}
\qquad
1&=&\sum_{\Gamma}\lambda_{i;\Gamma}^{*}\lambda_{i;\Gamma}
\langle \hat{m}_{i;\Gamma}\rangle_{\Psi_{0}}\;,\\
\langle \hcd_{i,\sigma}\hc_{i,\sigma'} \rangle_{\Psi_{0}}
&=&
\sum_{\Gamma}\lambda_{i;\Gamma}^{*}\lambda_{i;\Gamma}
\langle\hcd_{i,\sigma}\hc_{i,\sigma'} \hat{m}_{i;\Gamma}\rangle_{\Psi_{0}}\;,
\end{eqnarray}
\esup

With the expectation value 
\begin{equation}\label{34556}
m_{i;\Gamma,\Gamma'}=\langle\,  \ulm_{i;\Gamma,\Gamma'}  \rangle_{\Psi_{\rm G}}=
\langle \hat{m}_{i;\Gamma,\Gamma'}\rangle_{\Psi_{0}}
\lambda_{i;\Gamma}^{*}\lambda_{i;\Gamma'}
\end{equation}
where 
\begin{equation}
\langle \hat{m}_{i;\Gamma,\Gamma'}\rangle_{\Psi_{0}}=
\sum_{I}T_{i;I,\Gamma}T^{*}_{i;I,\Gamma'} m^0_{i;I}
\end{equation}
one can readily calculate the local energy 
 $\langle  \hat{H}_{i,{\rm loc}} \rangle_{\Psi_{\rm G}}$. 
For the calculation of the single-particle energy, we have to determine 
 the expectation values of normal and anomalous hopping operators, 
\begin{equation}\label{8.410bb} 
\big \langle  \hat{c}_{i,\sigma_1}^{(\dagger)}\hat{c}_{j,\sigma_2}^{(\dagger)} \big \rangle_{\Psi_{\rm G}}
=\sum_{\sigma'_1,\sigma'_2}\left(q_{\sigma_1}^{\sigma'_1}\right)^{(*)}
\left( q_{\sigma_2}^{\sigma'_2}\right)^{*}\big \langle  
\hat{c}_{i,\sigma'_1}^{(\dagger)}\hat{c}_{j,\sigma'_2}^{(\dagger)} \big \rangle_{\Psi_{0}}\;,
\end{equation}
where, to simplify the notation, we dropped the lattice site index of the  
`renormalisation matrix' $q_{\sigma}^{\sigma'}$. 
 This renormalisation matrix is given as
\begin{equation}\label{8.430a} 
q_{\sigma}^{\sigma'}=\sum_{\Gamma,\Gamma'}\lambda^{*}_{\Gamma}
\lambda^{}_{\Gamma'}\langle \Gamma|\hat{c}^{\dagger}_{\sigma}  
|\Gamma' \rangle
\sum_{I,I'}T^{\phantom{*}}_{I,\Gamma}T^{*}_{I',\Gamma'}
\langle \hat{H}^{\sigma'}_{I,I'}\rangle_{\Psi_0} \;,
\end{equation}
where we introduced the operator 
\begin{eqnarray}\label{9.700ab} 
\hat{H}^{\sigma'}_{I,I'}&\equiv&(1-f_{\sigma',I})\langle I'  |\hc_{\sigma'} |I'\cup \sigma'  \rangle
\hat{m}_{I,I'\cup \sigma'}\\\nonumber
&&\!\!\!\!\!\!\!\!\!\!\!\!\!\!\!\!\!\!\!\!
+\langle I \backslash \sigma' |\hc_{\sigma'} |I  \rangle
\left(
f_{\sigma',I'}\hat{m}_{I\backslash \sigma',I'}+
(1-f_{\sigma',I'})\hat{m}^{\sigma'}_{I\backslash \sigma',I'}
\right)\;.
\end{eqnarray}
 Here, we use the abbreviation $f_{\sigma,I}=\langle I|\hcd_{\sigma} 
\hc_{\sigma} |I \rangle $ and the operator 
\begin{equation}\label{9.710} 
\hat{m}^{\sigma}_{I,I'}\equiv \hat{C}^{\dagger}_{I}\hat{C}_{I'}
\prod_{\sigma'\in J\backslash \sigma }(1-\hat{n}_{\sigma'})\;,
\end{equation}
which is defined for $\sigma\in J\equiv \overline{I\cup I'}$. 
 Note that the expectation value of (\ref{9.700ab}) in Eq. (\ref{8.430a})
can be readily evaluated with equations~(\ref{78965}).
   
\section{The Kotliar-Ruckenstein Theory}\label{chap3}
In the first part of this section, we introduce  the auxiliary  particle 
 method, which was proposed by Kotliar and Ruckenstein for an investigation
 of the single-band Hubbard model. In the second part,
 previous attempts to generalise this approach for 
 multi-band Hubbard models are discussed.
\subsection{The One-Band Model}\label{chap3a}
We start from the Hamiltonian~(\ref{5.20}) with its 
 four-dimensional local 
Hilbert space $\mathcal{H}_{i}$ for each lattice site $i$, represented by the 
 four states $\ket{I}=\ket{\emptyset}$, $\ket{\sigma}$ ,$\ket{12}$ 
(with $\sigma=1,2$ for the two spin directions). 
 The Hilbert space of the whole lattice system is given by the 
 tensor product 
\begin{equation}
\mathcal{H}\equiv\underset{\scriptstyle  i}{\otimes}\; \mathcal{H}_{i} \;.
\end{equation}

Kotliar and Ruckenstein  
 introduced auxiliary bosonic operators $\hat{\phi}^{\dagger}_{i;I}$, 
 $\hat{\phi}^{}_{i;I}$,  which lead to an enlarged local Hilbert space
 $\mathcal{H}^{\rm FB}_{i}$
  defined by the basis states 
\bsup\label{9.90g}
 \begin{equation}\label{9.90}
\ket{I,I'}_{i;{\rm FB}}\equiv \ket{I}_{i}\otimes \ket{I'}_{i;{\rm B}}\;.
\end{equation}
Here, $\ket{I}_i$ is the fermionic configuration state, defined in 
equation~(\ref{4.30a}), and $\ket{I}_{i;{\rm B}}$ is the bosonic state
\begin{equation}\label{9.100}
\ket{I}_{i;{\rm B}}\equiv \hat{\phi}^{\dagger}_{i;I}\ket{0}_{i;{\rm B}}\;.
\end{equation}
\esup
 with the bosonic vacuum state $\ket{0}_{i;{\rm B}}$. 
The original quantum mechanical problem can be recovered in the following 
 way:\newline
i) One has  to find a subspace $\ulHcal_{i}$ of 
$\mathcal{H}^{\rm FB}_{i}$ which is isomorphic 
to the physical Hilbert space~$\mathcal{H}_{i}$ for each lattice site $i$.  
Kotliar and Ruckenstein defined this subspace by means of the constraints 
\bsup\label{9.110g}
\begin{eqnarray}\label{9.110a}
\qquad
\hat{F}_{i,0}&\equiv&1-\sum_{I}\hat{n}^{\rm B}_{i;I}=0\;,\\\label{9.110b}
\hat{F}_{i,\sigma}&\equiv&\hcd_{i,\sigma}\hc_{i,\sigma}
-\hat{n}^{\rm B}_{i;12}-\hat{n}^{\rm B}_{i;\sigma}=0 \;,
\end{eqnarray}
\esup
with the bosonic occupation operators
\begin{equation}\label{9.115}
\hat{n}^{\rm B}_{i;I}\equiv \hat{\phi}^{\dagger}_{i;I}\hat{\phi}^{}_{i;I}\;.
\end{equation} 
The constraints (\ref{9.110g}) define the subspace $\ulHcal_{i}$ via the conditions
\begin{equation}
\hat{F}_{i,\tilde{\sigma}}\ket{\ulPsi}=0
\end{equation}
 for each $\ket{\ulPsi}\in \ulHcal_{i}$ and $\tilde{\sigma}\in(0,1,2$).
Alternatively, we can define $\ulHcal_{i}$  directly by specifying 
 its basis
\begin{equation}\label{9.120}
\ket{\ulI}_{i}\equiv
\ket{I,I}_{i;{\rm FB}}=\ket{I}_i{\otimes}\ket{I}_{i;{\rm B}}\;.
\end{equation}
The corresponding Hilbert space for the lattice system is given by
 \begin{equation}
\ulHcal\equiv\underset{\scriptstyle  i}{\otimes}\; \ulHcal_{i} \;.
\end{equation}
Note that, by construction, there is now a 
one-to-one correspondence  of all physical states 
$\ket{\Psi}\in \mathcal{H}$ and their counterparts 
 $\ket{\ulPsi}\in\ulHcal $.\newline
ii) With the auxiliary Hilbert spaces $\ulHcal_{i}$ 
and $\ulHcal$ properly defined,
 one can find  operators $\ulO_i$ in  
$\ulHcal_{i}$ that are {\sl similar}
 to the physical operators $\hat{O}_i$ in $\mathcal{H}_{i}$.
Here, `similarity' means that 
\begin{equation}\label{9.130}
\langind{i} \ulI| \ulO_i |  \ulI'\rangind{i}=
\langind{i} I| \hat{O}_i |  I'\rangind{i}
\end{equation} 
for all configurations $\ket{I}$, $\ket{I'}$. 
With similar local operators  $\ulO_i$,
 one can set up an `effective' Hamiltonian $\ulH_{1{\rm B}}$ which is similar 
to the physical Hamiltonian $\hat{H}_{1{\rm B}}$, i.e., it obeys 
\begin{equation}
\langle \Psi| \hat{H}_{1{\rm B}}| \Psi'\rangle=
\langle \ulPsi| \ulH_{1{\rm B}} |\ulPsi'\rangle
\end{equation}
for all physical states $\ket{\Psi},\ket{\Psi'}\in \mathcal{H}$ and their 
 counterparts $\ket{\ulPsi},\ket{\ulPsi'}\in\ulHcal$.
 In this way, we have introduced an exact mapping of the
 original physical problem, described by the Hamiltonian 
$\hat{H}_{1{\rm B}}$ in its Hilbert space $\mathcal{H}$
  and the effective Hamiltonian $\ulH_{1{\rm B}}$ in $\ulHcal$ .

To set up $\ulH_{1{\rm B}}$,  we start with an identification of 
operators that are 
 similar to the  fermionic 
operators $\hat{c}^{(\dagger)}_{i,\sigma}$ 
in $\mathcal{H}_{i}$.
 Their counterparts $\ulc^{(\dagger)}_{i,\sigma}$ in 
$\ulHcal_{i}$ can be chosen as
\bsup\label{9.140g}
  \begin{equation}\label{9.140}
\qquad
\ulc^{\dagger}_{i,\sigma}
=\hat{r}^{}_{i,\sigma}\hat{c}^{\dagger}_{i,\sigma}\;\;\;\;,\;\;\;\;
\ulc^{}_{i,\sigma}
=\hat{r}^{\dagger}_{i,\sigma}\hat{c}^{}_{i,\sigma}\;,
\end{equation}
where the bosonic operators
\begin{eqnarray}\label{9.150a}
\qquad
\hat{r}^{}_{i,\sigma}&\equiv&
\hat{\phi}^{\dagger}_{i;12}\hat{\phi}^{}_{i;\bar{\sigma}}
+\hat{\phi}^{\dagger}_{i;\sigma}\hat{\phi}^{}_{i;\emptyset}\;,\\
\hat{r}^{\dagger}_{i,\sigma}&=&\hat{\phi}^{\dagger}_{i;\bar{\sigma}}
\hat{\phi}^{}_{i;12}
+\hat{\phi}^{\dagger}_{i;\emptyset}\hat{\phi}^{}_{i;\sigma}\;.
\end{eqnarray}
\esup
have been introduced. As required, the operators $\ulc^{(\dagger)}_{i,\sigma}$
 obey equation~(\ref{9.130}). To set up the Hamiltonian $\ulH_{1{\rm B}}$
 in $\ulHcal_{i}$, we further need to find an 
operator $\ulm_{i;12}$  
that is similar to $\hat{m}_{i;12}=\hat{n}_{i,1}\hat{n}_{i,2}$.
The most obvious choice is 
\begin{equation}\label{9.160}
\ulm_{i;12}=\hat{n}^{\rm B}_{i;12}\;.
\end{equation} 
Note, however, that there is a large amount of arbitrariness. 
For example, the operators 
 \begin{equation}\label{9.170}
\ulm_{i;12}=\hat{n}^{\rm B}_{i;I} 
\hat{m}_{i;12}
\;\;\;\;{\rm or} \;\;\;\;\ulm_{i;12}=\hat{m}_{i;12}
\end{equation} 
are also similar to $\hat{m}_{i;12}$ since both obey equation~(\ref{9.130}).
 The same ambiguity arises for the operators~(\ref{9.140}). 
 For example, they  may equally well be chosen as 
\bsup\label{9.180g}
 \begin{equation}\label{9.175}
\ulc^{\dagger}_{i,\sigma}
=\hat{q}^{}_{i,\sigma}\hat{c}^{\dagger}_{i,\sigma}\;\;\;\;,\;\;\;\;
\ulc^{}_{i,\sigma}
=\hat{q}^{\dagger}_{i,\sigma}\hat{c}^{}_{i,\sigma}
\end{equation}
with
\begin{eqnarray}\label{9.180a}
\qquad
\hat{q}^{}_{i,\sigma}
&\equiv& (\hat{\Delta}_{i,\sigma})^{-1/2} \hat{r}^{}_{i,\sigma}
(1-\hat{\Delta}_{i,\sigma})^{-1/2}\;,\\
\hat{q}^{\dagger}_{i,\sigma}
&=&(1-\hat{\Delta}_{i,\sigma})^{-1/2} 
\hat{r}^{\dagger}_{i,\sigma}(\hat{\Delta}_{i,\sigma})^{-1/2}\;,
\end{eqnarray}
and
\begin{equation}\label{9.190}
\hat{\Delta}_{i,\sigma}\equiv
\hat{n}^{\rm B}_{i;12}+\hat{n}^{\rm B}_{i;\sigma}\;.
\end{equation}
\esup
In fact, this choice is better
 than~(\ref{9.140}) and was used by Kotliar and Ruckenstein since 
 it yields the correct 
 ground-state energy in the uncorrelated limit $U=0$ 
  if the resulting effective Hamiltonian
\begin{equation}\label{9.200}
\qquad
\ulH_{1{\rm B}}=\sum_{i,j,\sigma} 
t_{i,j} \hat{q}^{}_{i,\sigma} \hcd_{i,\sigma}  
\hat{q}^{\dagger}_{j,\sigma} \hc_{j,\sigma}+
U\sum_i\hat{n}^{\rm B}_{i;12} 
\end{equation}
is investigated on a mean-field level, see below. 

Kotliar and Ruckenstein \cite{kotliar1986} used 
 a functional integral approach to calculate the free energy 
 of the Hamiltonian (\ref{9.200}). For ground-state properties, i.e., at
 zero temperature, their saddle-point approach is  equivalent to a replacement 
 of the bosonic operators $\hat{\phi}^{}_{i;I}$ by the amplitudes 
 $\varphi_{i;I}$. They govern the bosonic occupations
\begin{equation}
\langle \hat{n}^{\rm B}_{i;I}  \rangle  = n^{\rm B}_{i;I}=|\varphi_{i;I}|^2\;.
\end{equation}
 and have to be  determined
 by a minimisation of the ground-state energy functional
\begin{equation}\label{9.240}
\langle  
 \ulH_{1{\rm B}}\rangle_{\Psi^{\rm FB}_0}
=\sum_{i,j,s} 
t_{i,j} q^{*}_{i,\sigma}q^{}_{j,s}   \langle \hcd_{i,\sigma}  
\hc_{j,s}\rangle_{\Psi_0}+
U\sum_i n^{\rm B}_{i;12}\;.
\end{equation}
Here, the factors $q^{*}_{i,\sigma}$ and $q^{}_{i,\sigma}$ are defined in 
(\ref{9.180g}) with the 
 operators $\hat{\phi}^{(\dagger)}_{i;I}$ replaced by $\varphi^{(*)}_{i;I}$. 
Note 
that all quantities in this  section are real and the asterisks, e.g., 
 in equation~(\ref{9.240}), are only 
 used in anticipation of the corresponding multi-band results in 
section~\ref{chap5}.

Instead of dealing with the 
 exact constraints~(\ref{9.110g}),
 they are also satisfied only on a `mean-field level' by Kotliar and 
Ruckenstein, i.e., for the expectation values 
\bsup\label{9.260g}
\begin{eqnarray}\label{9.260a}
\qquad
1&=&\sum_{I}n^{\rm B}_{i;I}\;,\\
n^0_{i,\sigma}&=&n^{\rm B}_{i;12}+n^{\rm B}_{i;\sigma}\;.
\end{eqnarray}
\esup 

The constraints~(\ref{9.260g}) and the energy functional~(\ref{9.240}) 
 are the same as those derived for the Gutzwiller wave function 
 in the limit of infinite spatial dimensions \cite{metzner1988,gebhard1987b} or evaluated 
 by means of the Gutzwiller approximation \cite{gutzwiller1963,vollhardt1984,buenemann1998b}.

\subsection{Multi-Band Hubbard Models}\label{chap3b} 
A generalisation of the slave-boson theory is straightforward
\cite{hasegawa1997,fresard1997} for
 multi-band Hubbard models with a local Coulomb interaction of the form
\begin{equation}\label{9.270}
\hat{H}_{\rm I}=\sum_{\sigma,\sigma'}U_{\sigma,\sigma'}\hat{n}_{\sigma}
\hat{n}_{\sigma'}=\sum_{I}U_{I}\hat{m}_I\;,
\end{equation}
where 
\begin{equation}
U_{I}=\sum_{\sigma,\sigma'\in I}U_{\sigma,\sigma'}\;
\end{equation}
It yields the same energy functional  as derived 
 within the Gutzwiller theory \cite{buenemann1997c}. 
For the treatment of 
 general multi-band Hubbard models a generalised  slave-boson theory 
has been derived by Dai et al.\ 
 \cite{dai2006} and, more successfully, by Lechermann et al. \cite{lechermann2007}.
 
 As demonstrated in the previous sections, the slave-boson approach contains 
  a number of adjustable objects. 
 These rather flexible elements of the theory are the definition of the 
extended Hilbert space $\mathcal{H}^{\rm FB}_i$, the definition of its physical 
 subspace $\ulHcal_i$, the form of the constraint 
equations $\hat{F}_{i,\sigma}=0$ 
 and, finally, the particular definition of similar operators  $\ulO_i$.
 Despite this huge flexibility, in both works \cite{dai2006} and  
\cite{lechermann2007}, the authors fail to 
 derive an exact mapping of Hilbert spaces and Hamiltonians which, 
 on a mean-field level, leads to satisfactory results.
We will briefly summarise the previous attempts to formulate a 
 generalised slave-boson scheme for multi-band Hubbard models in this section. 
 Our own derivation for such an approach is discussed
 in section \ref{chap5}.

 Dai et al.\ use constraint equations which
  do not define the correct physical Hilbert space. This 
 problem has been pointed out and solved by  Lechermann et al. 
In their work, however,  
they fail to derive proper fermionic 
operators $\ulc^{(\dagger)}_{i}$. Instead, 
 symmetry arguments are used in order to guess the form of
 certain operators $\ulc^{(\dagger)}_{i}$, 
  which lead to a reasonable energy functional on mean-field level.
 These operators, however,
 are  {\sl not} similar to the  physical operators   $\hat{c}^{(\dagger)}_{i}$, 
 see below.
Therefore,  the whole derivation seems even less controlled than for the
 single-band model.
In this section, we briefly summarise the main ideas of 
 the slave-boson mean-field theory  introduced by 
 Lechermann et al.
  
As in case of the single-band model 
 one has to set up a local Hamiltonian $\ulHcal_i$ which is isomorphic to the 
 physical fermionic Hamiltonian $\mathcal{H}_i$ 
with its basis $\ket{\Gamma}_i$.
Lechermann et al.\ discuss
 various possibilities to define such local Hilbert spaces consisting of
 bosons and fermions. At first sight, it seems that 
$\ulHcal_i$ is most naturally defined 
   as a generalisation of~(\ref{9.120}) through a basis
\begin{equation}\label{9.340}
\ket{\ulGamma}_i\equiv\ket{\Gamma}_i\otimes\ket{\Gamma}_{i;\rm B}
\end{equation}
with bosonic operators $\hat{\phi}^{(\dagger)}_{i;\Gamma}$
 and the corresponding states 
$\ket{\Gamma}_{i;\rm B}\equiv \hat{\phi}^{\dagger}_{i;\Gamma}
\ket{0}_{i;\rm B}$. However, 
 the Hilbert-space 
 $\ulHcal_i$  defined by this  basis 
is discarded
by Lechermann et al.\ on the grounds that 
  there is no way to find a reasonable set 
  of constraint equations as an alternative definition of~$\ulHcal_i$. 
In section~\ref{chap5}, we 
 show that the basis~(\ref{9.340}) can, in fact, be used for a 
 slave-boson theory, which, however, has to be different  from 
 the original scheme introduced by Kotliar and Ruckenstein.

Instead of~(\ref{9.340}), Lechermann et al.\ introduce the basis
\begin{equation}\label{9.350}
\ket{\ulGamma}_i\equiv\frac{1}{\sqrt{|\Gamma|}}
\sum_{I(|I|=|\Gamma|)} \hat{\phi}^{\dagger}_{i;\Gamma,I}\ket{0}_{i;\rm B}  \otimes \ket{I}_i
\end{equation}
 as a definition of their Hilbert space $\ulHcal_i$. 
Note that in their derivation they draw a distinction
 between physical particles and quasi-particles, described by operators 
$\hat{d}^{(\dagger)}_{i,\sigma}$ and $\hat{f}^{(\dagger)}_{i,\sigma}$, 
respectively. Our derivation in this section indicates that 
this distinction is unnecessary.

  The Ansatz~(\ref{9.350}) employs bosonic operators 
  $\hat{\phi}_{i;\Gamma,I}$ 
 for each pair of multiplet states $\ket{\Gamma}$ and 
 configurations states $\ket{I}$ with the same particle number
 $|\Gamma|=|I|$.   
 The number of these operators is much larger than the 
 dimension of the local Hilbert space, which is 
 different from the original single-band scheme 
introduced by Kotliar and Ruckenstein. 
 Despite this large number of operators, 
the space  $\ulHcal_i$ is isomorphic to the physical 
 Hilbert space $\mathcal{H}_i$  with its basis $\ket{\Gamma}_i$. 
 Therefore, it is possible to find operators $\ulO_i$ in  
$\ulHcal_{i}$ that are {\sl similar}
 to the physical operators $\hat{O}_i$ in $\mathcal{H}_{i}$, see below. 

For the mean-field treatment, one has to find constraints which
 define the Hilbert space $\ulHcal_{i}$ in a unique way. 
As shown in 
 \cite{lechermann2007}, this is achieved by means of the operator identities
\bsup\label{9.360g}
 \begin{eqnarray}\label{9.360a}
\qquad
 \hat{F}_{i,0}&=&1-\sum_{\Gamma,I}  \hat{\phi}^{\dagger}_{i;\Gamma,I} \hat{\phi}^{}_{i;\Gamma,I}=0\;,\\\label{9.360b}
\hat{F}_{i;\sigma,\sigma'}&=&\hcd_{i,\sigma}\hc_{i,\sigma'}\\\nonumber
&&-
\sum_{\Gamma,I,I'}\hat{\phi}^{\dagger}_{i;\Gamma,I'} \hat{\phi}^{}_{i;\Gamma,I}
\,\langind{i} I | \hcd_{i,\sigma}\hc_{i,\sigma'}   |I'\rangind{i}=0\;.
\end{eqnarray}
\esup

The representation $\ulm_{i;\Gamma,\Gamma'}$ of local operators  in $\ulHcal_i$
 is readily given by 
\label{9.370}
\begin{equation}
\ulm_{i;\Gamma,\Gamma'}=\sum_{I}
\hat{\phi}^{\dagger}_{i;\Gamma,I} \hat{\phi}^{}_{i;\Gamma',I}\;.
\end{equation}
This result leads to the representation
\begin{equation}\label{9.380}
\ulH_{i;\rm loc}=\sum_{\Gamma,\Gamma'}
E^{\rm loc}_{i;\Gamma,\Gamma'} 
\sum_{I}
\hat{\phi}^{\dagger}_{i;\Gamma,I} \hat{\phi}^{}_{i;\Gamma',I}
\end{equation}
of the local Hamiltonian~(\ref{4.10a}). 
 
In order to set up the effective Hamiltonian $\ulH$,  
one needs  representations $\ulc^{\dagger}_{i,\sigma}$ of fermionic 
creation operators. 
 As shown in \cite{lechermann2007}, a conceivable choice for 
$\ulc^{\dagger}_{i,\sigma}$ would be
\bsup\label{9.390g}
\begin{equation}\label{9.390}
\ulc^{\dagger}_{i,\sigma}=\sum_{\sigma'}
\hat{q}^{\sigma'}_{i,\sigma}\hcd_{i,\sigma'}\;,
\end{equation}
where
\begin{equation}\label{9.400}
 \hat{q}^{\sigma'}_{i,\sigma}=\sum_{\Gamma,\Gamma'}\sum_{I,I'}
\frac{\langind{i} \Gamma|\hcd_{i,\sigma} |\Gamma'\rangind{i}
\langind{i} I|\hcd_{i,\sigma} |I'\rangind{i}}
{\sqrt{|\Gamma|(N-|\Gamma'|)}}
\hat{\phi}^{\dagger}_{i;\Gamma,I} \hat{\phi}^{}_{i;\Gamma',I'}
\end{equation}
\esup
is a bosonic operator and~$N$ is the number of spin-orbital 
states $\ket{\sigma}$ per site.
 Evaluated on mean-field level, however, expression~(\ref{9.390g}) does not lead 
to a reasonable energy functional since it does not yield
 the correct results in the uncorrelated limit. Note that the situation here
 is different from the single-band model since, there,
 it was possible to find improved expressions for the operator 
  $\ulc^{\dagger}_{i,\sigma}$, which are still similar to the 
 physical operators $\hcd_{i,\sigma}$. 
In case of the operators~(\ref{9.390g}),
 it seems to be	unfeasible to improve them accordingly. Instead, 
 Lechermann et al.\ introduce the following `improved' expression for~(\ref{9.390g}), 
which, though leading to reasonable results on the mean-field level, 
 is mathematically {\sl not} similar to  $\hcd_{i,\sigma}$, 
\begin{eqnarray}\label{9.410}
\qquad
 \hat{q}^{\sigma'}_{i,\sigma}&=&
\sum_{\Gamma,\Gamma'}\sum_{I,I'}\sum_{\gamma}
\langind{i} \Gamma|\hcd_{i,\sigma} |\Gamma'\rangind{i}
\langind{i} I|\hcd_{i,\gamma} |I'\rangind{i}\\
&&\;\;\;\;\;\;\;\;\;\;\;\;\;\;\;\;\;\;\;\;\;
\times \hat{\phi}^{\dagger}_{i;\Gamma,I} \hat{\phi}^{}_{i;\Gamma',I'}
\hat{M}_{i;\gamma,\sigma'}
\;,
\end{eqnarray}
where
\begin{equation}\label{9.420}
\hat{M}_{i;\sigma,\sigma'}\equiv
\left(\frac{1}{2}
[\hat{\Delta}_{i}^{\rm (p)}\hat{\Delta}_{i}^{\rm (h)}
+\hat{\Delta_{i}^{\rm (h)}}\hat{\Delta_{i}^{\rm (p)}}]^{-1/2} 
\right)_{\sigma,\sigma'}\;,
\end{equation}
and 
\bsup\label{9.430g}
\begin{eqnarray}\label{9.430a}
\qquad
\hat{\Delta}_{i;\sigma,\sigma'}^{\rm (p)}&=&\sum_{\Gamma,I,I'}
\hat{\phi}^{\dagger}_{i;\Gamma,I} \hat{\phi}^{}_{i;\Gamma,I'}
\langind{i} I'|\hcd_{i,\sigma} \hc_{i,\sigma'}    | I \rangind{i}
\;,\\
\hat{\Delta}_{i;\sigma,\sigma'}^{\rm (h)}&=&\sum_{\Gamma,I,I'}
\hat{\phi}^{\dagger}_{i;\Gamma,I} \hat{\phi}^{}_{i;\Gamma,I'}
\langind{i} I'|\hc_{i,\sigma'} \hcd_{i,\sigma}    | I \rangind{i}
\;.
\end{eqnarray}
\esup
Note  that $\hat{\Delta}_{i;\sigma,\sigma'}^{\rm (h/p)}$ 
 and $\hat{M}_{i;\sigma,\sigma'}$ are 
 considered as matrices with respect to the indices $\sigma,\sigma'$
  whose elements are bosonic operators. The inversion $[\ldots]^{-1/2}$ 
and the square root
in (\ref{9.420}) are 
defined with respect to  this matrix structure. 

The operators~(\ref{9.380}) and~(\ref{9.410}) define an effective Hamiltonian 
\begin{equation}\label{9.440}
\ulH=
\sum_{i,j}t^{\sigma,\sigma'}_{i,j}\sum_{\sigma,\sigma'}
\sum_{\gamma,\gamma'}\hat{q}^{\gamma}_{i,\sigma} 
 \left(\hat{q}^{\gamma'}_{j,\sigma'} \right)^{\dagger}
\hcd_{i,\gamma}
\hc_{j,\gamma'} 
+\sum_i\ulH_{i;{\rm loc}}
\end{equation}
which can now be evaluated  on mean-field level, i.e., 
  by replacing the operators 
$\hat{\phi}^{(\dagger)}_{i;\Gamma,I}$ by their corresponding amplitudes
$\varphi^{(*)}_{i;\Gamma,I}$. 
 These amplitudes then serve as variational parameters. 

In Ref.~\cite{buenemann2007c}, it was shown that 
 the energy functional that 
 results from the mean-field treatment of the 
constraints~(\ref{9.360g}) and of the Hamiltonian~(\ref{9.440}) 
agrees with the Gutzwiller variational results introduced in 
 Refs.~\cite{buenemann1998,buenemann2005}.

\section{A new Slave-Boson Theory for Multi-Band Hubbard Models}\label{chap5}
As discussed in the previous section, a generalisation 
of the Kotliar-Ruckenstein scheme for the investigation 
 of multi-band models faces significant problems, which, 
up to now, have not been solved satisfactorily.
Here, we show that, due to the enormous flexibility 
 of the slave-boson approach, it is, in fact, relatively easy 
 to reproduce the Gutzwiller energy functional for multi-band Hubbard models.
 To this end, however, one has to approach the problem in a  
 different way than Kotliar and Ruckenstein. In the first part of this
 section, we introduce our new slave-boson scheme by 
  reconsidering the single-band model. In the second part, 
we show that  our new approach can be easily applied to multi-band models
 including those with superconducting ground states. 
  
\subsection{The Single-Band Model}\label{chap5a}
For our alternative formulation of the slave-boson theory, 
we introduce the operators 
\bsup\label{4598}
\begin{equation}\label{4598a}
m^{\rm B}_{i;I}= \hat{\theta}^{\dagger}_{i;I}
\hat{\theta}^{}_{i;I}
\end{equation}
where
\begin{equation}
\hat{\theta}^{\dagger}_{i;I}\equiv\hat{\phi}^{\dagger}_{i;I}
\prod_{I'} \hat{e}_{i;I'}
\end{equation}
and 
\begin{equation}
\qquad
\hat{e}_{i;I}\equiv\prod_{n=1}^{\infty}\bigg(1-\frac{\hat{\phi}^{\dagger}_{i;I}
\hat{\phi}^{}_{i;I}}
{n}\bigg)
\end{equation}
\esup
 is the projection operator onto the vacuum state of the boson
  created by $\hat{\phi}^{\dagger}_{i;I}$. The operator (\ref{4598a})
 therefore projects onto the sub-space with exactly one boson in the state 
  $|i;I \rangle $ occupied.

As pointed out before, there is a large amount of arbitrariness 
in the choice, e.g., of the constraints~(\ref{9.110g}). 
 Instead of those equations, we can also work with
\bsup
\label{9.500g}
\begin{eqnarray}\label{9.500a}
\qquad
\hat{F}_{i,0}&\equiv&1-\sum_{I}\hat{m}_{i;I}m^{\rm B}_{i;I}=0
\;,\\\label{9.500b}
\hat{F}_{i,\sigma}&\equiv&\hat{n}_{i,\sigma}
-\hat{n}_{i,\sigma}\sum_{I}\hat{m}_{i;I}  m^{\rm B}_{i;I}=0\;,
\end{eqnarray}
\esup
where
 $\hat{m}_{i;I}$ has been defined in Eq.~(\ref{4.52a}). 
 Note that the Hilbert space $\ulH_i$, given by the basis~(\ref{9.120}),
 is already uniquely defined by the first constraint, equation~(\ref{9.500a}).
 In $\ulH_i$, however, the second equation is 
 equally valid, i.e., we have $\hat{F}_{i,\sigma}\ket{\ulI}_i=0$ for all states~(\ref{9.120}). 

In the Kotliar-Ruckenstein scheme,  the operators $\ulm_{i;I}$ are chosen as 
  $n^{\rm B}_{i;I}$, c.f. equation (\ref{9.115}). 
In our approach, we work with
\begin{equation}\label{9.510}
\ulm_{i;I}\equiv\hat{m}_{i;I} m^{\rm B}_{i;I}    \;.
\end{equation}

Since we are only interested in the derivation of an approximate ground 
state-energy functional, we avoid functional integral techniques here and 
employ the variational wave function
 \bsup\label{9.210g}
\begin{equation}\label{9.210}
\ket{\Psi^{\rm FB}_0}\equiv\ket{\Psi^{\rm B}_{0}}\otimes\ket{\Psi_{0}}
\end{equation}
where 
$\ket{\Psi_{0}}$ is a fermionic single-particle product state 
and 
\begin{equation}\label{9.220a}
\ket{\Psi^{\rm B}_{0}}\equiv \prod_{i}\hat{D}_{i}\ket{0}
\end{equation}
a coherent bosonic state with 
\begin{equation}\label{9.220b}
\hat{D}_i\equiv 
\prod_{I}\exp{\left(\varphi_{i;I}\hat{\phi}^{\dagger}_{i;I}-
 \varphi^{*}_{i;I}\hat{\phi}_{i;I}\right)}\;.
\end{equation}
\esup
By construction,~(\ref{9.220a}) is a normalised
 eigenstate of 
$\hat{\phi}^{}_{i;I}$ with eigenvalues $\varphi_{i;I}$; 
see, e.g., Ref.~\cite{negele1988}. In addition, we have
\bsup
\begin{eqnarray}
\qquad
\langle \hat{e}_{i;I}  \rangle_{\Psi^{\rm B}_{0}}
&=&\exp{(- |\varphi_{i;I}|^2)}\;,\\
\langle   \theta^{\dagger}_{i,I}  \theta^{}_{i,I}    \rangle_{\Psi^{\rm B}_{0}}
&=& |\varphi_{i;I}|^2\prod_{I'} \exp{(- |\varphi_{i;I'}|^2)}
\end{eqnarray}
\esup
which leads to the expectation value
\begin{equation}
\langle  \ulm_{i;I}  \rangle_{\Psi^{\rm FB}_{0}} = 
\langle\hat{m}_{i;I}\rangle_{\Psi_0} 
|\vartheta_{i;I}|^2
\end{equation}
where we introduced
\begin{equation}
 \vartheta_{i;I} \equiv  \varphi_{i;I}  \prod_{I'}\exp{(-|\varphi_{i;I'}|^2/2)}\;.
\end{equation}
 A comparison with the corresponding result in the Gutz\-willer theory
 \cite{buenemann1998} shows that  the Gutzwiller 
variational  parameters $\lambda_I$ correspond to the 'renomalised' 
bosonic amplitudes
\begin{equation}\label{9.530}
 \vartheta_{i;I} \hat{=}  \lambda_{i;I} \;.
\end{equation} 
 
The constraints~(\ref{9.500g}), evaluated on mean-field level 
(i.e., using the wave-functions (\ref{9.210}) ),
 have the form
\bsup\label{9.540g}
\begin{eqnarray}\label{9.540a}
\qquad
1&=&\sum_{I}|\vartheta_{i;I}|^2\langle\hat{m}_{i;I}
\rangle_{\Psi_0}\;,
\\\label{9.540b}
n^0_{i,\sigma}&=&|\vartheta_{i;\sigma}|^2
\langle\hat{m}_{i;\sigma}\rangle_{\Psi_0}
+|\vartheta_{i;12}|^2
\langle\hat{m}_{i;12}\rangle_{\Psi_0}
\end{eqnarray}
\esup
which is in agreement with equations (\ref{9.260g})
 if we equate $n^{\rm B}_{i;I}$ (in the Kotliar-Ruckenstein scheme)
  with 
$|\vartheta_{i;I}|^2\langle\hat{m}_{i;I}\rangle_{\Psi_0}$ 
(in our new slave-boson scheme).

Finally, we choose the operators $\ulc^{\dagger}_{i;\sigma}$ in $\ulH_i$ as 
\bsup
  \begin{equation}\label{9.550}
\ulc^{\dagger}_{i,\sigma}
=\hat{q}^{}_{i,\sigma}\hat{c}^{\dagger}_{i,\sigma}\;\;\;\;,\;\;\;\;
\ulc^{}_{i,\sigma}
=\hat{q}^{\dagger}_{i,\sigma}\hat{c}^{}_{i,\sigma};\,
\end{equation}
where
\begin{equation}\label{9.560}
\hat{q}^{}_{i,\sigma}\equiv
\hat{\theta}^{\dagger}_{i;12}\hat{\theta}^{}_{i;\bar{\sigma}}\hat{n}_{i,\bar{\sigma}}
+\hat{\theta}^{\dagger}_{i;\sigma}\hat{\theta}^{}_{i;\emptyset}
(1-\hat{n}_{i,\bar{\sigma}})\;.
\end{equation}
\esup
Note that here, unlike in the Kotliar-Ruckenstein scheme,
 the operators $\hat{q}^{(\dagger)}_{i,\sigma}$ contain
both fermionic and bosonic degrees of freedom. Evaluated with the wave 
function~(\ref{9.210}), one finds
\begin{equation}\label{9.570}
q^{}_{i,\sigma}=\langle \hat{q}^{}_{i,\sigma} \rangle_{\Psi^{\rm FB}_0}=
\vartheta^{*}_{i;12}\vartheta^{}_{i;\bar{\sigma}}n^0_{i,\bar{\sigma}}
+\vartheta^{*}_{i;\sigma}\vartheta^{}_{i;\emptyset}
(1-n^0_{i,\bar{\sigma}})\;,
\end{equation}
 which agrees with equations~(\ref{9.180g}) on mean-field level. However, 
 the expectation value of a hopping operator is the same as
in the Gutzwiller theory,
\begin{equation}\label{9.580}
\langle \ulc^{\dagger}_{i,\sigma}\ulc^{}_{j,\sigma'} \rangle_{\Psi^{\rm FB}_0}
=q^{}_{i,\sigma}q^{*}_{j,\sigma}\langle 
\hcd_{i,\sigma}\hc_{j,\sigma}
\rangle_{\Psi_0}\;,
\end{equation}
only if we neglect the `three-line' contributions 
\begin{equation}\label{9.590}
(\vartheta^{*}_{i;12}\vartheta^{}_{i;\bar{\sigma}}-
\vartheta^{*}_{i;\sigma}\vartheta^{}_{i;\emptyset})^2
|\langle \hcd_{i,\bar{\sigma}}\hc_{j,\bar{\sigma}} \rangle_{\Psi_0}|^2
\langle \hcd_{i,\sigma}\hc_{j,\sigma} \rangle_{\Psi_0}\approx 0\;.
\end{equation}
These terms emerge when the fermionic expectation value 
$\langle \hat{n}_{i,\bar{\sigma}} \hcd_{i,\sigma}\hc_{j,\sigma}  \hat{n}_{j,\bar{\sigma}} \rangle_{\Psi_{0}}$ 
is evaluated by means of Wick's theorem,
\begin{eqnarray}\label{9.600}
\langle \hat{n}_{i,\bar{\sigma}} \hcd_{i,\sigma}\hc_{j,\sigma}  \hat{n}_{j,\bar{\sigma}} \rangle_{\Psi_{0}}
&=&n^0_{i,\bar{\sigma}}n^0_{j,\bar{\sigma}}\langle \hcd_{i,\sigma}\hc_{j,\sigma} \rangle_{\Psi_0}\\\nonumber
&&+|\langle \hcd_{i,\bar{\sigma}}\hc_{j,\bar{\sigma}} \rangle_{\Psi_0}|^2
\langle \hcd_{i,\sigma}\hc_{j,\sigma} \rangle_{\Psi_0}\;.
\end{eqnarray}
Since
  the three-line  terms vanish in the limit of  infinite spatial dimensions, 
 our slave-boson approach  yields the same variational  
ground-state energy as the Gutz\-willer theory. 

\subsection{General Multi-Band Models}\label{chap5b}
A generalisation of our new slave-boson scheme for multi-band Hubbard models 
 is straightforward. We work with an arbitrary set of local 
 multiplet states $\ket{\Gamma}_i$, which define the basis 
(\ref{9.340}) of a Hilbert space $\ulHcal_i$. For these states, 
 we introduce bosonic operators  $\hat{m}^{\rm B}_{i;\Gamma}$ and  
$\hat{\theta}^{(\dagger)}_{i;\Gamma}$ as in  (\ref{4598}) only
 with $I$ replaced by~$\Gamma$.

 As a generalisation of~(\ref{9.500g}), we work with the 
 constraints  
\bsup \label{9.610g}
\begin{eqnarray}\label{9.610a}
\qquad
\hat{F}_{i,0}&\equiv&1-\sum_{\Gamma}\hat{m}_{i;\Gamma}\hat{m}^{\rm B}_{i;\Gamma}=0\;,\\\label{9.610b}
\hat{F}_{i;\sigma,\sigma'}&\equiv&
\hcd_{i,\sigma}\hc_{i,\sigma'}-
\hcd_{i,\sigma}\hc_{i,\sigma'}
\sum_{\Gamma}\hat{m}_{i;\Gamma}\hat{m}^{\rm B}_{i;\Gamma}=0\;,
\end{eqnarray}
\esup
which yield an alternative way to define the Hilbert space
  $\ulHcal_i$.

The operators $\hat{m}_{i;\Gamma,\Gamma'}$ are properly represented in 
$\ulHcal_i$ by \begin{equation}\label{9.620}
  \ulm_{i;\Gamma,\Gamma'}\equiv \hat{m}_{i;\Gamma,\Gamma'}
\hat{\theta}^{\dagger}_{i;\Gamma}\hat{\theta}^{}_{i;\Gamma'}\;.
\end{equation}
 An evaluation of these operators on mean-field level, i.e, by means of a 
wave function~(\ref{9.210g}), with 
\begin{equation}\label{9.630}
\hat{D}_{i}\equiv \prod_{\Gamma}
\exp{(\varphi_{i;\Gamma}\hat{\phi}^{\dagger}_{i;\Gamma}-
 \varphi^*_{i;\Gamma}\hat{\phi}^{}_{i;\Gamma}  )}\;,
\end{equation}
leads to
\begin{equation}\label{9.640}
m_{i;\Gamma,\Gamma'}=\langle\,  \ulm_{i;\Gamma,\Gamma'}  \rangle_{\Psi^{\rm FB}_{0}}=
\langle \hat{m}_{i;\Gamma,\Gamma'}\rangle_{\Psi_{0}}
\vartheta_{i;\Gamma}^{*}\vartheta_{i;\Gamma'}
\end{equation}
with 
\begin{equation}
\vartheta_{i;\Gamma}\equiv \varphi_{i;\Gamma}
\prod_{\Gamma'}\exp{(-|\varphi_{i;\Gamma'}|^2/2)} \;.
\end{equation}
A comparison with Eq. (\ref{34556}) reveals the correspondence
 of the variational parameters $\lambda_{i;\Gamma}$ in the 
 Gutzwiller theory and the amplitudes  $\vartheta_{i;\Gamma}$ 
in our new slave-boson mean-field approach. 

An evaluation of the constraints~(\ref{9.610g}) 
on mean-field  level leads to
\bsup\label{9.670g}
\begin{eqnarray}
\qquad
1&=&\sum_{\Gamma}\vartheta_{i;\Gamma}^{*}\vartheta_{i;\Gamma}
\langle \hat{m}_{i;\Gamma}\rangle_{\Psi_{0}}\;,\\
\langle \hcd_{i,\sigma}\hc_{i,\sigma'} \rangle_{\Psi_{0}}
&=&
\sum_{\Gamma}\vartheta_{i;\Gamma}^{*}\vartheta_{i;\Gamma}
\langle\hcd_{i,\sigma}\hc_{i,\sigma'} \hat{m}_{i;\Gamma}\rangle_{\Psi_{0}}\;,
\end{eqnarray}
 \esup
which matches the Gutzwiller constraints, 
 equations~(\ref{2786}). 

Finally, we define the operator 
\bsup\label{9.680g}
\begin{equation}\label{9.680}
\ulc^{\dagger}_{i,\sigma}\equiv 
\sum_{\sigma'}\hat{q}_{i,\sigma}^{\sigma'} \hcd_{i,\sigma'}
\end{equation}
with 
\begin{equation}\label{9.690}
\hat{q}_{\sigma}^{\sigma'}=
\sum_{\Gamma,\Gamma'}\hat{\theta}^{\dagger}_{\Gamma}
\hat{\theta}^{}_{\Gamma'}
\langle\Gamma | \hcd_{\sigma} |\Gamma'\rangle
\sum_{I,I'}T^{\phantom{*}}_{I,\Gamma}T^{*}_{I',\Gamma'}
\hat{H}^{\sigma'}_{I,I'}\;.
\end{equation}
\esup
Here, we dropped the lattice-site index $i$ and use the operator 
 $\hat{H}^{\sigma'}_{I,I'}$ defined in Eq. (\ref{9.700ab}).   
 The operator $\ulc^{\dagger}_{i,\sigma}$ in $\ulHcal_i$ 
is similar to the physical creation operator $\hcd_{i,\sigma}$ 
because the sums over $\sigma'$ and $I,I'$
in equations~(\ref{9.680g}) are just a complicated 
 expression for
\begin{equation}
\ket{\Gamma}\bra{\Gamma'}=
\sum_{\sigma'}\sum_{I,I'}T^{\phantom{*}}_{I,\Gamma}T^{*}_{I',\Gamma'}
\hat{H}^{\sigma'}_{I,I'}\hcd_{\sigma'}\;.
\end{equation}

A mean-field evaluation of~(\ref{9.690}) leads to the renormalisation 
 matrix
\begin{eqnarray}\label{9.720}
\qquad
q_{\sigma}^{\sigma'}&=&
\sum_{\Gamma,\Gamma'}\vartheta^{*}_{i;\Gamma}
\vartheta^{}_{i;\Gamma'}
\langle\Gamma | \hcd_{i,\sigma} |\Gamma'\rangle\\\nonumber
&&\times \sum_{I,I'}T^{\phantom{*}}_{I,\Gamma}T^{*}_{I',\Gamma'}
\langle \hat{H}^{\sigma'}_{I,I'}\rangle_{\Psi_0}\;,
\end{eqnarray}
which matches equation~(\ref{8.430a}) in the Gutzwiller theory. 
As in the single-band model, the expectation 
 values of normal and anomalous hopping operators agree with those in the Gutzwiller
 theory
\begin{eqnarray}  \nonumber
\qquad
&&\langle \ulc^{(\dagger)}_{i,\sigma_1} \ulc^{(\dagger)}_{j,\sigma_2} \rangle_{\Psi^{\rm FB}_{0}}\\
&&=\sum_{\sigma_1',\sigma_2'}\left(q_{i,\sigma_1}^{\sigma_1'}\right)^{(*)}
\left(q_{j,\sigma_2}^{\sigma_2'}\right)^{(*)}
\big \langle  \hat{c}^{(\dagger)}_{i,\sigma_1'} 
\hat{c}^{(\dagger)}_{j,\sigma_2'} \big  \rangle_{\Psi_{0}} \;,
\end{eqnarray}
only if we neglect the contributions with more than one line 
 connecting the sites $i$ and $j$. This is ensured in 
 the limit of infinite spatial dimensions,  where both 
 approaches then yield the same ground-state energy functional.

\section{Conclusions}\label{chap6}
In summary, we have developed a new slave-boson scheme for general 
 multi-band Hubbard models which, in the limit of infinite 
 dimensions, reproduces the results of the Gutz\-willer theory. 
 The main advantage of our new approach is its exactness 
 up to the point where a mean-field approximation is applied. In addition,
 it automatically covers the cases of systems with superconducting order 
 parameters. 
      
  \bibliographystyle{unsrt}
\bibliography{bib3}
\end{document}